\documentclass[12pt]{article}
\usepackage[dvips]{graphicx}
\usepackage{amsmath}
\usepackage{graphicx}
\usepackage{amsfonts}
\usepackage{amssymb}
\usepackage{epsfig}
\usepackage{subfigure}
\usepackage{xcolor}
\usepackage{tikz}
\usepackage[colorlinks=blue]{hyperref}
\definecolor{purple}{rgb}{1,0,1}
\definecolor{lime}{HTML}{A6CE39} 

\definecolor{lime}{HTML}{A6CE39}
\newcommand{\orcidicon}{%
	\begin{tikzpicture}
	\draw[lime, fill=lime] (0,0) 
		circle [radius=0.16] 
		node[white] {{\fontfamily{qag}\selectfont \tiny ID}};
	\draw[white, fill=white] (-0.0625,0.095) 
		circle [radius=0.007];
	\end{tikzpicture}
	\hspace{-5mm}
}
\newcommand\orcidPetarpa{{\href{https://orcid.org/0000-0003-0357-9620}{\orcidicon}}}
\newcommand\orcidMatt{{\href{https://orcid.org/0000-0003-1088-6485}{\orcidicon}}}

\begin{document}
\title{\bf Near-horizon geodesics for astrophysical\\ and idealised black holes:\\
Coordinate velocity and coordinate acceleration}
\author{Petarpa Boonserm,$^{1}$\orcidPetarpa{}\hspace{1mm} Tritos Ngampitipan,$^{2}$ and Matt Visser$^{3}$\orcidMatt{}\\
$^{1}$  {\small {\em Department of Mathematics and Computer Science, Faculty of Science,}}\\
{\small {\em Chulalongkorn University, Bangkok 10330, Thailand}}\\
$^{2}$  {\small {\em Faculty of Science, Chandrakasem Rajabhat University,}}\\
{\small {\em Bangkok 10900, Thailand}}\\
$^{3}$  {\small {\em School of Mathematics and Statistics, Victoria University of Wellington,}}\\
{\small {\em PO Box 600, Wellington 6140, New Zealand}}\\
}
\date{16 October 2017; 28 May 2018; \LaTeX-ed \today}
\maketitle
\def\d{{\mathrm{d}}}
\newcommand{\red}[1]{{\color{red} #1}}
\newcommand{\blue}[1]{{\color{blue} #1}}
\section*{Abstract}
Geodesics (by definition) have an intrinsic 4-acceleration zero. However, when expressed in terms of coordinates, the \emph{coordinate acceleration} $\d^2 x^i/\d t^2$ can very easily be non-zero, and the \emph{coordinate velocity} $\d x^i/\d t$ can behave unexpectedly. The situation becomes extremely delicate in the near-horizon limit{---for both astrophysical and idealised black holes}---where an inappropriate choice of coordinates can quite easily lead to significant confusion. We shall carefully explore the relative merits of horizon-penetrating \emph{versus} horizon-non-penetrating coordinates, arguing that in the near-horizon limit the coordinate acceleration $\d^2 x^i/\d t^2$
is best interpreted in terms of horizon-penetrating coordinates.

\bigskip
\noindent\textsc{Pacs}: 04.20.-q;  04.70.-s; 04.70.Bw; 02.40.Hw

\bigskip
\noindent\textsc{Keywords}: Geodesic equation, coordinate velocity, coordinate acceleration, horizon-penetrating coordinates, horizon-non-penetrating coordinates.

\bigskip
\noindent\textsc{Published}:  Universe {\bf 4} (2018) 68
\clearpage
\hrule

\tableofcontents

\bigskip
\hrule
\section{Introduction}
\parindent0pt
\parskip8pt

Coordinate dependence in general relativity is a topic that continues to cause confusion to this day, despite over 100 years of work on this issue. (For a variety of articles, both pro and con, both~published and unpublished, see  \cite{McGruder:2017, Thiago:2015, Felber:2009, Felber:2005, Mashhoon:2005, Krori:1985,Spallicci:2017, Spallicci:2014,Abbott:2017a, Abbott:2017b, Celerier:2016, Felber:2011, Loinger:2009,Ohanian:2011}. For two recent overviews, see~\cite{Iorio:2015, Debono:2016}). The~situation is particularly acute in the immediate vicinity of any horizon that might be present, {whether it be for an astrophysical or an idealised (mathematical) black hole,} where~an inappropriate choice of coordinates can needlessly add to the confusion. Indeed, while~horizons are often associated with coordinate singularities, these coordinate singularities are a property of the coordinate patch, not the spacetime geometry, and these coordinate singularities can quite easily go away with a different choice of coordinates.
{For astrophysical black holes, as opposed to maximally analytically extended idealised black holes, one still trusts the usual Einstein equations in the domain of outer communication---and down to any inner horizon that might be present. Similarly for the black holes arising from numerical simulations, which are key to modelling the astrophysical black holes of direct observational interest, one typically calculates down to some region inside the outer horizon, but well above the singular region, relying on the usual idealised picture for near-(outer)-horizon physics. Finally, for semi-classical black holes, as long as the quantum fields are in the Unruh vacuum state, the near-horizon geometry in the vicinity of the future horizon is qualitatively similar to that in classical general relativity.

In short, the black holes of observational interest in astronomy and cosmology can be adequately represented,  
at least in the domain of outer communication and down to any inner horizon that might be present, by the idealised Schwarzschild and Kerr spacetimes---and analysis of the near-horizon physics can adequately be performed using the classical Schwarzschild and Kerr spacetimes. 
}

In fact it is very useful to distinguish:
\begin{itemize}
\item Horizon-penetrating coordinates --- these coordinate systems are regular as one crosses the horizon (for example, Painleve--Gullstrand coordinates, Kerr--Schild coordinates, and variants thereof).
\item
Horizon-non-penetrating coordinates --- these coordinate systems are singular as one crosses the horizon (for example, the Schwarzschild curvature coordinates, isotropic coordinates, and variants thereof).
\end{itemize}
The horizon-non-penetrating coordinates are simpler for some purposes, (the metric is typically diagonal), but are ill-behaved in the immediate vicinity of the horizon. In contrast horizon-penetrating coordinates are better behaved in the immediate vicinity of the horizon,
but the metric is typically non-diagonal, and the asymptotic behaviour may sometimes be more subtle than expected.
We shall work through a number of examples illustrating the dangers and the pitfalls.

Consider for instance the Schwarzschild geometry --- this is a very well-known spacetime since it was the first known exact solution to the (vacuum) Einstein field equations~\cite{Schwarzschild}. It is certainly of direct physical relevance --- the spacetime geometry exterior to the sun can be well-approximated by the Schwarzschild geometry. Perhaps the simplest form of the Schwarzschild spacetime is the Hilbert form expressed in terms of (what are now known as)  Schwarzschild curvature coordinates  \cite{Hilbert1,Hilbert2,Hilbert3}
\begin{equation}
\d s^{2} = -\left(1 - \frac{2m}{r}\right)\d t^{2} + \left(1 - \frac{2m}{r}\right)^{-1}
\d r^{2} + r^{2}\left(\d\theta^{2} + \sin^{2}\theta\, \d\phi^{2}\right).
\end{equation}
There is a coordinate singularity at $r=2m$, see for instance~\cite{Wheeler,Wald,Hobson,Stephani,Parry:2012,Muller:2009,Visser:1995,regular,Lake:1994,Czerniawski:2002}, making this representation horizon-non-penetrating~\cite{Fromholz:2013}.  It is easy to see that in these coordinates the radial geodesics ``pile up'' at $r=2m$, never (in these coordinates) crossing the horizon. In fact, for any radial incoming geodesic, $\dot r \to 0$ as one approaches the horizon.

Taking into account the Killing conservation law for the energy, we shall soon see that, even for infalling particles, $\ddot r \to \hbox{(something positive)}$ sufficiently close to the horizon, though not at the horizon itself. However, this near-horizon $\ddot r >0$ phenomenon is a coordinate artefact; the behaviour can be very different in other coordinates. Despite this, some researchers are now (even in 2018) completely misinterpreting this coordinate artefact and asserting  that ``gravity becomes repulsive near the horizon.'' This claim is, at best, a gross misinterpretation of the actual situation. (For specific examples of this particular confusion, see particularly  \cite{McGruder:2017, Felber:2009, Felber:2005, Mashhoon:2005,Krori:1985,Abbott:2017a,Abbott:2017b,Celerier:2016,Felber:2011,Loinger:2009}. For partial antidotes, see  \cite{Spallicci:2017, Spallicci:2014, Ohanian:2011}. For a somewhat different sort of coordinate confusion, mistaking white holes for black holes, see  \cite{Thiago:2015}.)

Below, we shall show that the coordinate acceleration near horizons is, in horizon-penetrating coordinates, (such as the Painleve--Gullstrand~\cite{Painleve,Gullstrand,unexpected,Acoustic,Analogue,river,heuristic,Finch:2012,Rosquist:2003} or  Kerr--Schild~\cite{Wheeler,Stephani,Visser:1995,regular,Lake:1994,Rosquist:2003} coordinates), much easier to understand. We shall then wrap up with some generic comments regarding arbitrary horizon-penetrating coordinate systems~\cite{Wiltshire:2009,Kerr,Rosquist:2009,Newman-Janis}.

We shall use letters from the beginning of the Roman alphabet ($a$, $b$, $c$, $d$, ...) for spacetime indices, (see for instance Wald~\cite{Wald}, or Hobson--Efstathiou--Lasenby~\cite{Hobson}). Whenever there is a clearly defined time coordinate $t$, we  shall use letters from the middle of the Roman alphabet ($i$, $j$, $k$, $l$, ...) for the remaining spatial indices.
We reserve the notation $\dot x$ and $\ddot x$ for derivatives with respect to the time coordinate $t$. 

\section{Geodesic equation}
Consider the geodesic equation in non-affine-parameterized form:
\begin{equation}
\frac{d^{2}x^{a}}{d\lambda^{2}} + \Gamma^{a}{}_{bc}\;\frac{dx^{a}}{d\lambda}\frac{dx^{b}}{d\lambda} = f(\lambda)\frac{dx^{a}}{d\lambda}.
\end{equation}
This part of the analysis works equally well for timelike or null geodesics.
Assume the zero'th coordinate is timelike, at least outside any horizon that might be present. That is, take $x^{a} = (t, x^{i})$.  We can then choose the coordinate $t$ to be a non-affine parameter for the geodesic. The geodesic equation separates into:
\begin{eqnarray}
\frac{d^{2}t}{dt^{2}} + \Gamma^{t}{}_{bc}\;\frac{dx^{b}}{dt}\frac{dx^{c}}{dt}     &=& f(t)\,\frac{dt}{dt} \\[5pt]
\frac{d^{2}x^{i}}{dt^{2}} + \Gamma^{i}{}_{bc}\;\frac{dx^{b}}{dt}\frac{dx^{c}}{dt} &=& f(t)\,\frac{dx^{i}}{dt}.
\end{eqnarray}
The first of these equations implies
\begin{equation}
f(t) = \Gamma^{t}{}_{bc}\frac{dx^{b}}{dt}\frac{dx^{c}}{dt}.
\end{equation}
The second equation then becomes
\begin{equation}
\frac{d^{2}x^{i}}{dt^{2}} = -\Gamma^{i}{}_{bc}\; \frac{dx^{b}}{dt}\frac{dx^{c}}{dt} + \left(\Gamma^{t}{}_{bc}\; \frac{dx^{b}}{dt}\frac{dx^{c}}{dt}\right)\frac{dx^{i}}{dt}.
\end{equation}
This is still very general.
Let us now specialize to spherical symmetry, taking
\begin{equation}
g_{ab} = \left[ \begin{array}{cc|cc}
g_{tt} & g_{tr} & 0 & 0 \\ g_{tr} & g_{rr} &0&0\\
\hline
0&0&g_{\theta\theta}&0\\ 0 &0&0& g_{\phi\phi}
\end{array}\right].
\end{equation}
Then the radial geodesics are given by
\begin{equation}
\frac{d^{2}r}{dt^{2}} = -\Gamma^{r}{}_{bc}\frac{dx^{b}}{dt}\frac{dx^{c}}{dt} + \left(\Gamma^{t}{}_{bc}\frac{dx^{b}}{dt}\frac{dx^{c}}{dt}\right)\frac{dr}{dt}.
\end{equation}
That is,
\begin{eqnarray}
\ddot{r} &=& -\left[\Gamma^{r}{}_{tt} + 2\Gamma^{r}{}_{rt}\,\dot{r} + \Gamma^{r}{}_{rr}\, \dot{r}^{2}\right] + \left[\Gamma^{t}{}_{tt} + 2\Gamma^{t}{}_{rt}\dot{r} + \Gamma^{t}{}_{rr}\,\dot{r}^{2}\right]\dot{r}.
\label{rdotdot}
\end{eqnarray}
Finally, regrouping, we see
\begin{eqnarray}
\ddot{r} &=& -\Gamma^{r}{}_{tt} + \left(\Gamma^{t}{}_{tt} - 2\Gamma^{r}{}_{rt}\right)\,\dot{r} + \left(2\Gamma^{t}{}_{rt} - \Gamma^{r}{}_{rr}\right)\,\dot{r}^{2} + \Gamma^{t}{}_{rr}\;\dot{r}^{3}.\label{rdotdot2}
\end{eqnarray}
Note that the ``coordinate acceleration'' $\ddot r$ is \emph{cubic} in the ``coordinate velocity'' $\dot r$.  This effect is certainly real if perhaps naively unexpected. (This effect is also manifestly coordinate dependent.)

\section[Killing conservation law for energy: Coordinate velocity]
{Killing conservation law for energy: \\ Coordinate velocity}

In all the situations we will be interested in there is a timelike Killing vector, (timelike outside any horizon that may be present), and there is no real loss of generality in taking the $t$ coordinate to be compatible with that Killing vector; so that $K^a = (\partial_t)^a$. (That is, we choose coordinates to \emph{manifestly} respect the time-translation Killing symmetry.) But then any timelike geodesic with 4-velocity $V^a$ is subject to the energy conservation law
\begin{equation}
g_{ab} \; K^a V^b = -\epsilon,
\label{E:killing}
\end{equation}
where $\epsilon$ is a constant (effectively the energy per unit rest mass; $\epsilon=1$ corresponds to dropping a particle at rest from spatial infinity; $\epsilon>1$ corresponds to dropping a moving particle from spatial infinity; $\epsilon<1$ corresponds to a gravitationally bound particle, dropped at rest from some finite radius).\footnote{In fact, in an asymptotically flat spacetime,  $\epsilon = {1\over\sqrt{1-\beta_\infty^2}}$, where $\beta_\infty$ is the ``coordinate velocity at infinity''.}
In spherical symmetry this Killing conservation law can be written as
\begin{equation}
g_{tb} \;  {(1,\dot r,0,0)^b\over||(1,\dot r,0,0)||} = -\epsilon.
\end{equation}
That is
\begin{equation}
(g_{tt} + g_{tr}\, \dot r) = -\epsilon \sqrt{ -( g_{tt}+2 g_{tr} \,\dot r + g_{rr} \,\dot r^2)}.
\label{E:condition-1}
\end{equation}
Even more explicitly
\begin{equation}
(g_{tt} + g_{tr}\, \dot r)^2 = - \epsilon^2 ( g_{tt}+2 g_{tr} \,\dot r + g_{rr} \,\dot r^2).
\label{E:condition-2}
\end{equation}
This is \emph{quadratic} in $\dot r$, with general solution
\begin{equation}
\dot r =  {-g_{tr}(1+ \epsilon^{-2}g_{tt})
\pm \sqrt{ (1+\epsilon^{-2}g_{tt})(g_{tr}^2-g_{tt}g_{rr})  }
\over g_{rr} +\epsilon^{-2} g_{tr}^2}.
\label{E:quadratic00}
\end{equation}

Physically the situation is this: If one drops a particle from some initial position $r_0$ with initial coordinate velocity $\dot r_0$, then one can calculate the energy $\epsilon$ from equation (\ref{E:killing}), and subsequently extract $\dot r$ at general positions $r$ from equation
(\ref{E:quadratic00}).

Formally, null geodesics can be viewed as the $\epsilon \to \infty$ limit of this formalism. This is most easily seen from equations (\ref{E:condition-1}) or (\ref{E:condition-2}) which in the $\epsilon \to \infty$ limit imply
\begin{equation}
 g_{tt}+2 g_{tr} \,\dot r + g_{rr} \,\dot r^2 = 0.
\end{equation}
However this is exactly the condition that the radial curve is a null curve. In this null limit one sees that the Killing conservation law becomes
\begin{equation}
\dot r =  {-g_{tr}
\pm \sqrt{g_{tr}^2-g_{tt}g_{rr}  }
\over g_{rr}}.
\label{E:quadratic-000}
\end{equation}

Two special cases are of particular interest:
\begin{itemize}
\item
In coordinate charts where the metric is diagonal, (for example, the Schwarzschild curvature coordinates or isotropic coordinates), we have $g_{tr}=0$. So for timelike geodesics:
\begin{equation}
\dot r =
\pm  {\sqrt{ (1+\epsilon^{-2}g_{tt})(-g_{tt}\,g_{rr})  }
\over \,g_{rr}}.
\label{E:quadratic0-a}
\end{equation}
As long as we are primarily interested in dropping (infalling) particles we must choose the negative root and set
\begin{equation}
\dot r =-  {\sqrt{ (1+\epsilon^{-2}g_{tt})(-g_{tt}\,g_{rr})  }
\over \,g_{rr}}.
\label{E:quadratic0-a2}
\end{equation}
In the null limit this simplifies considerably and becomes
\begin{equation}
\dot r =-  \sqrt{-g_{tt}\over g_{rr}  }.
\label{E:quadratic0-a3}
\end{equation}
\item
In contrast, in coordinate charts where the metric satisfies $g_{tt}g_{rr}-g_{tr}^2=-1$, (for example, the Painleve--Gullstrand coordinates or Kerr--Schild coordinates), for timelike geodesics we have:
\begin{equation}
\dot r
=  {-g_{tr}(1+\epsilon^{-2}g_{tt})
\pm  \sqrt{ 1+\epsilon^{-2}g_{tt}  }
\over  g_{rr} + \epsilon^{-2}g_{tr}^2}.
\label{E:quadratic0-b}
\end{equation}
As long as we are primarily interested in dropping (infalling) particles we can safely choose the negative root and set\footnote{Note that $1+\epsilon^{-2}g_{tt} \geq 0$ in order to keep $\dot r$ real, while $g_{rr} + \epsilon^{-2}g_{tr}^2>0$ always, so choosing the negative root selects the ingoing geodesic.}
\begin{equation}
\dot r = {-g_{tr}(1+\epsilon^{-2}g_{tt})
-  \sqrt{ 1+\epsilon^{-2}g_{tt}  }
\over  g_{rr} + \epsilon^{-2}g_{tr}^2}
= -   \sqrt{ 1+\epsilon^{-2}g_{tt}}  \left\{  1 + g_{tr}\sqrt{1+\epsilon^{-2}g_{tt}}\over
g_{rr} + \epsilon^{-2}g_{tr}^2 \right\}.
\label{E:quadratic0-b2}
\end{equation}
In the null limit this simplifies considerably and becomes
\begin{equation}
\dot r
= -    \left\{  1 + g_{tr}\over   g_{rr}  \right\}.
\label{E:quadratic0-b3}
\end{equation}
\end{itemize}
Let us now apply these quite general considerations to study the fixed-energy coordinate acceleration.

\section{Coordinate acceleration}

For a dropped (timelike trajectory) particle the coordinate acceleration at arbitrary radius is thus an interplay between the geodesic equation
\begin{eqnarray}
\ddot{r} &=& -\Gamma^{r}{}_{tt} + \left(\Gamma^{t}{}_{tt} - 2\Gamma^{r}{}_{rt}\right)\,\dot{r} + \left(2\Gamma^{t}{}_{rt} - \Gamma^{r}{}_{rr}\right)\,\dot{r}^{2} + \Gamma^{t}{}_{rr}\,\dot{r}^{3},\label{rdotdot3}
\end{eqnarray}
and the Killing-induced coordinate velocity equation
\begin{equation}
\dot r =  {-g_{tr}(1+ \epsilon^{-2}g_{tt})
\pm \sqrt{ (1+\epsilon^{-2}g_{tt})(g_{tr}^2-g_{tt}g_{rr})  }
\over g_{rr} +\epsilon^{-2} g_{tr}^2}.
\label{E:quadratic3}
\end{equation}
Combining these results we would get something of the general form
\begin{equation}
\ddot r = f(\epsilon, r)
\end{equation}
where $f(\epsilon, r)$ is some explicit \emph{but coordinate-dependent} function.\footnote{ We could always use the chain rule to write $\ddot r = {d \dot r\over d r}\; {d r\over dt} = {d \dot r\over d r}\; \dot r ={1\over2} {d(\dot r^2)\over d r}$. This serves as a consistency check, and side-steps the geodesic equation, but when doing so one looses information regarding the coordinate velocity dependence of the coordinate acceleration.}
We shall now give a few examples of this phenomenon, focussing particularly on near-horizon behaviour.

\section{Example:  Schwarzschild geometry}
The Schwarzschild spacetime geometry is perhaps the pre-eminent example of an exact solution in general relativity~\cite{Schwarzschild,Hilbert1,Hilbert2,Hilbert3,Wheeler,Stephani,Parry:2012,Muller:2009,Visser:1995}.
As specific examples of near-horizon behaviour for the coordinate velocity $\dot r$ and coordinate acceleration $\ddot r$, let us consider the Schwarzschild
spacetime in four commonly occurring coordinate systems~\cite{regular, Lake:1994, Czerniawski:2002,Painleve, Gullstrand,unexpected,Acoustic,Analogue}: Schwarzschild curvature coordinates, isotropic coordinates, Painleve--Gullstrand coordinates, and Kerr--Schild coordinates.
\subsection{Schwarzschild curvature coordinates}
The Schwarzschild geometry in Schwarzschild curvature coordinates is described by
\begin{equation}
\d s^{2} = -\left(1 - \frac{2m}{r}\right)\d t^{2} + \left(1 - \frac{2m}{r}\right)^{-1}\d r^{2} + r^{2}\left(\d\theta^{2} + \sin^{2}\theta \d\phi^{2}\right).
\end{equation}

It is easy to calculate the Christoffel symbols and to verify that the geodesic equation implies
\begin{equation}
\ddot{r} = -\frac{m(1 - 2m/r)}{r^{2}} + \left[\frac{3m}{r^2(1 - 2m/r)}\right]\,\dot{r}^{2}.
\end{equation}
This can be rewritten as:
\begin{equation}
\ddot{r} = -\frac{m}{r^{2}} \left\{ \left(1-{2m\over r}\right) - \frac{3\,\dot{r}^{2}}{(1 - 2m/r)} \right\}.
\end{equation}
This gives the coordinate acceleration $\ddot r$ in terms of the Newtonian value $-m/r^2$, modified by relativistic corrections --- due to both spacetime geometry and the local coordinate velocity.
Furthermore, this already demonstrates, (working in terms of $r$ and $\dot r$), that $\ddot r$ \emph{changes sign} at the critical coordinate velocities given by
\begin{equation}
(\dot r)_*^2 = {1\over3} \left(1-{2m\over r}\right)^2.
\end{equation}
At large $r$, (that is,  weak fields), this sign flip takes place at $\dot r^2 \approx 1/3$;  this is mildly relativistic but certainly not ultra-relativistic.\footnote{In fact this sign flip takes place for both ingoing and outgoing geodesics.}
Furthermore, from the Killing conservation equation we deduce
\begin{equation}
\dot r =
\pm \left(1-{2m\over r}\right) \sqrt{ 1 - \epsilon^{-2}\left(1-{2m\over r}\right)  }.
\label{E:quadratic3}
\end{equation}
In particular at the horizon $(\dot r)_H =0$, and at spatial infinity we see $\lim_{r\to\infty} \dot r =\sqrt{1-\epsilon^{-2}}$ for fixed $\epsilon$.
Combining these geodesic and Killing results
\begin{equation}
\ddot{r} = -\frac{m}{r^{2}}\left(1 - {2m\over r}\right) \left( 1 -  {3(\epsilon^2-1+2m/r)\over\epsilon^2}  \right).
\end{equation}
Note that (for fixed $\epsilon$) the coordinate acceleration $\ddot r$ goes through zero and changes \emph{sign}
at the critical values of $r$ given by\footnote{In fact this sign flip takes place for both ingoing and outgoing geodesics.}
\begin{equation}
r_* = {6m\over 3- 2\epsilon^2}; \qquad \hbox{and} \qquad r_*=2m.
\end{equation}
In particular at the horizon $(\ddot r)_H =0$ for fixed $\epsilon$.

For a time-like particle dropped at rest from spatial infinity ($\epsilon=1$) this simplifies to\footnote{Note that asymptotically $\dot r \to \sqrt{2m/r}$, and $\ddot r \to -m/r^2$, as expected from the Newtonian limit.}
\begin{equation}
\dot r =  -{ \left(1-{2m\over r}\right) \sqrt{2m\over r  }}; \qquad
\ddot{r} = -\frac{m}{r^{2}} \left(1-{2m\over r}\right)\left( 1 - {6m\over r}  \right);  \qquad r_* \in \{6m,2m\}.
\end{equation}
Oddly enough (in this particular coordinate system) the coordinate acceleration switches sign at $r_*=6m$, the location of the innermost stable circular orbit (ISCO);
this is a coincidence, not anything fundamental.

For a light-like particle  ($\epsilon\to\infty$) this simplifies to\footnote{Note that a photon can have non-trivial coordinate velocity and non-zero coordinate acceleration even if its physical speed is always exactly equal to $c$. This is one of the reasons that the concepts of coordinate velocity and coordinate acceleration must be used with care and discretion.
}
\begin{equation}
\dot r =  -\left(1-{2m\over r}\right); \qquad
\ddot{r} = +2\,\frac{m}{r^{2}} \left(1-{2m\over r}\right);  \qquad r_* =2m.
\end{equation}

Now these particular observations are not new, dating back (at least) to Hilbert in 1915 and the mid-1920s~\cite{Hilbert1,Hilbert2,Hilbert3}. (It must be emphasized that Hilbert's comments have subsequently been grossly misinterpreted by some of the later commentators on this topic.)\footnote{See particularly  \mbox{\cite{McGruder:2017, Felber:2009, Felber:2005, Mashhoon:2005,Krori:1985,Abbott:2017a,Abbott:2017b,Celerier:2016,Felber:2011,Loinger:2009}.} For partial antidotes, see  \cite{Spallicci:2017, Spallicci:2014, Ohanian:2011}. For a different sort of coordinate confusion (mistaking white holes for black holes) see  \cite{Thiago:2015}.}
What is new in the current discussion is that we will now put these issues into a wider context emphasising the extent to which these results are simply coordinate artefacts.

\begin{figure}[!h]
\centerline{\psfig{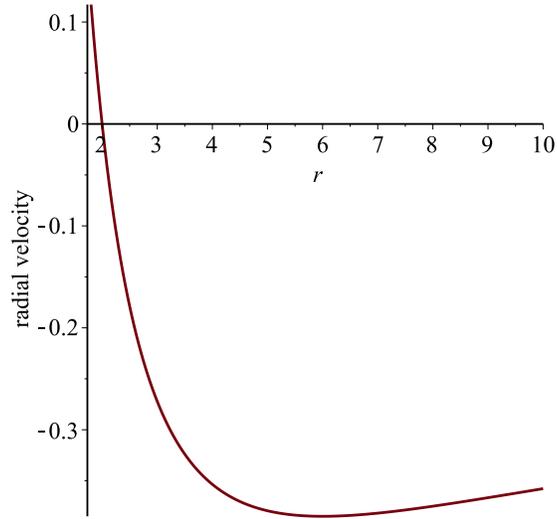}}
\vspace*{1pt}
\caption{Behaviour of $\dot{r}$ in the Schwarzschild geometry when using Schwarzschild  curvature coordinates for $m = 1$ and $\epsilon = 1$.
Note the curve crosses the $r$ axis only at $r=2$, and the coordinate velocity is negative all the way from the horizon to spatial infinity.}\label{F:sch-sch-dot-r}
\end{figure}

\begin{figure}[!h]
\centerline{\psfig{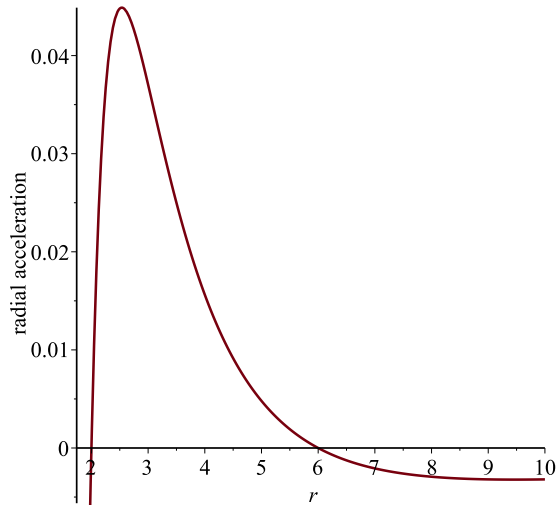}}
\vspace*{1pt}
\caption{Behaviour of $\ddot{r}$ in the Schwarzschild geometry when using Schwarzschild  curvature coordinates for $m = 1$ and $\epsilon = 1$.
Note the curve crosses the $r$ axis at both $r=2$ and $r=6$; the coordinate acceleration is positive between the horizon and the ISCO.}\label{F:sch-sch-ddot-r}
\end{figure}

\begin{figure}[!h]
\centerline{\psfig{file=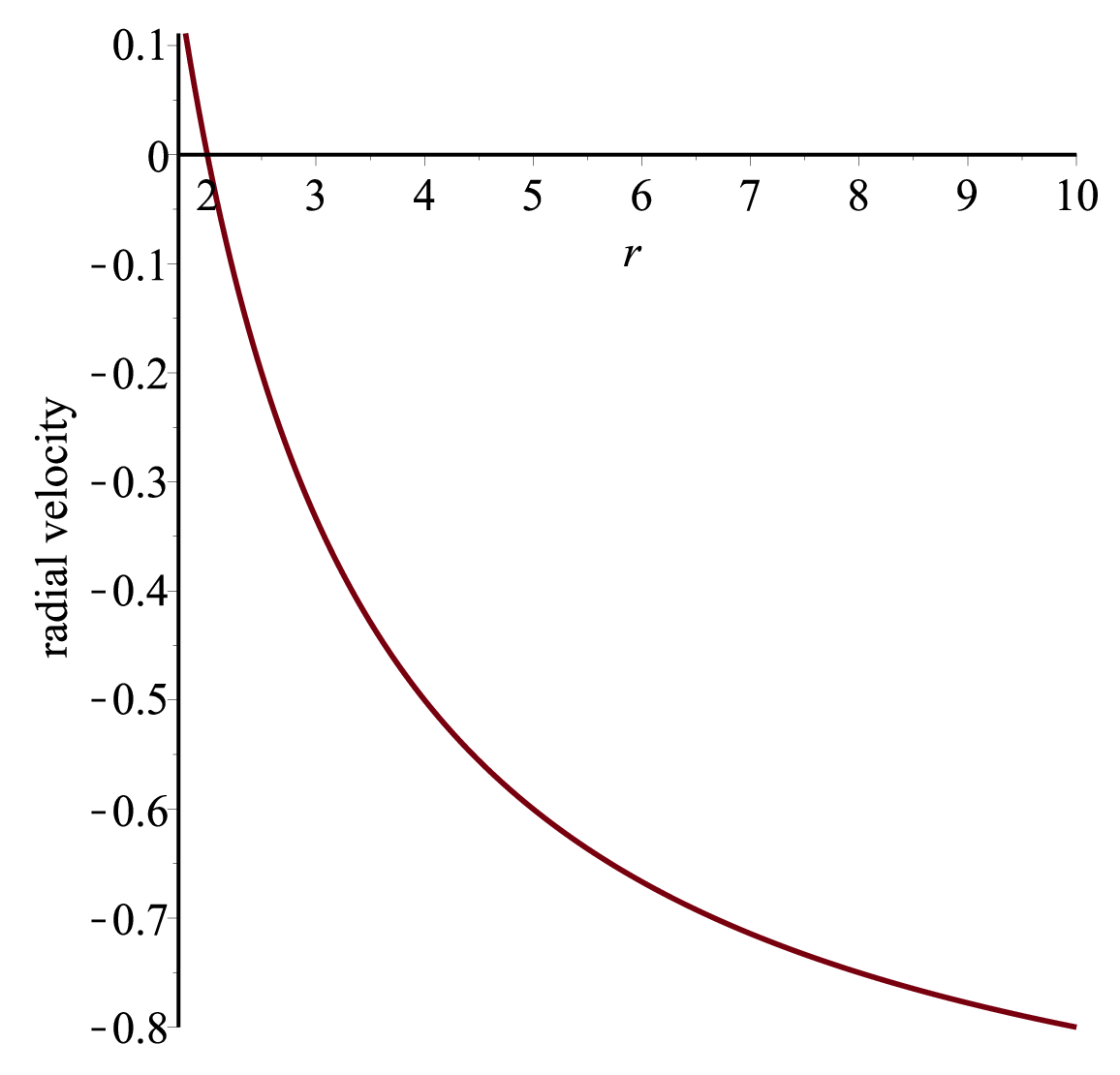,width=2.5in}}
\vspace*{1pt}
\caption{Behaviour of $\dot{r}$ for null geodesics in the Schwarzschild geometry when using Schwarzschild  curvature coordinates for $m = 1$ and $\epsilon\to\infty$.
Note the curve crosses the $r$ axis only at $r=2$, and the coordinate velocity is negative all the way from the horizon to spatial infinity.}\label{F:sch-sch-null-dot-r}
\end{figure}

\begin{figure}[!h]
\centerline{\psfig{file=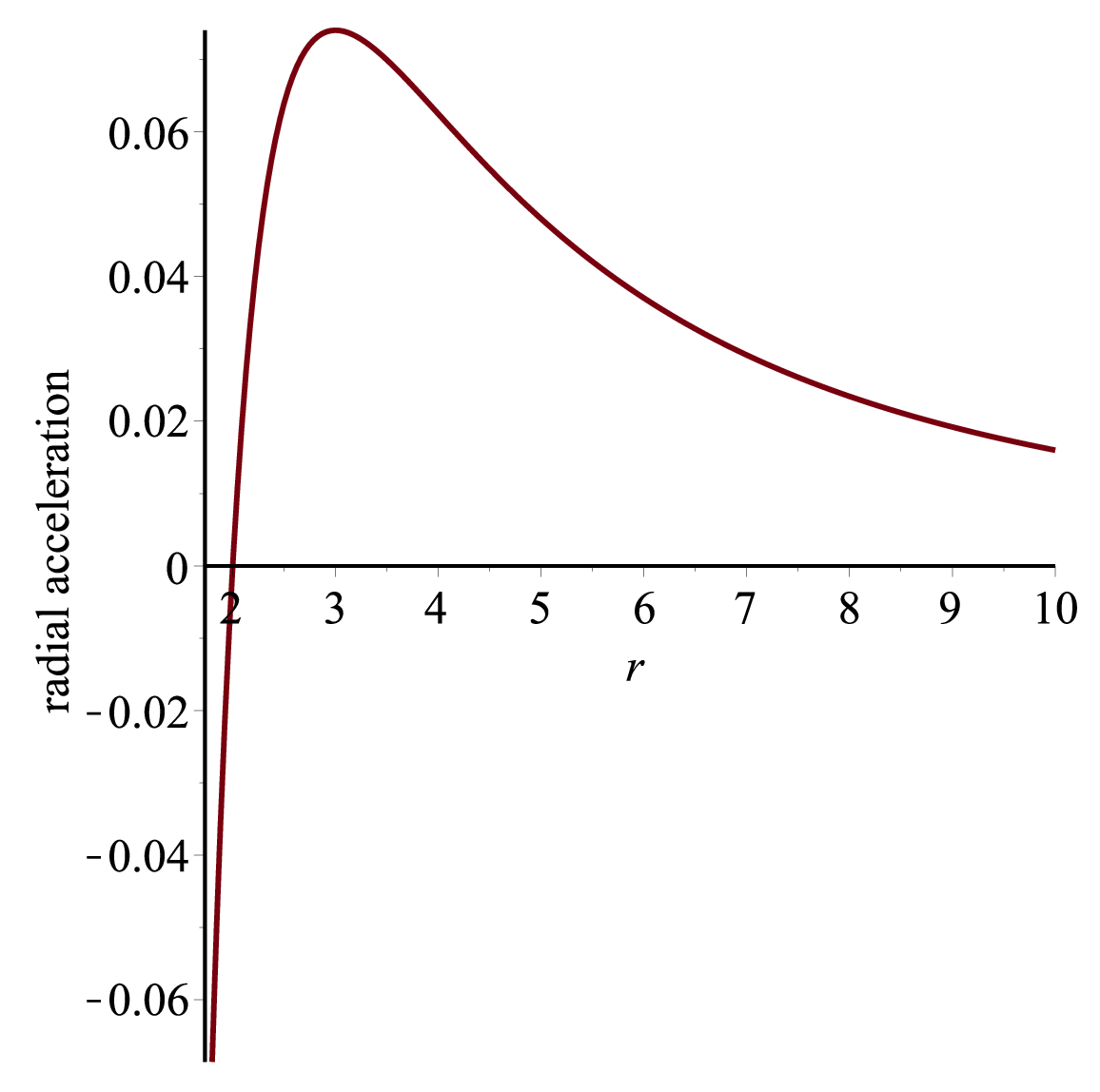,width=2.5in}}
\vspace*{1pt}
\caption{Behaviour of $\ddot{r}$ for null geodesics in the Schwarzschild geometry when using Schwarzschild  curvature coordinates for $m = 1$ and $\epsilon\to\infty$.
Note the curve crosses the $r$ axis only at $r=2$, and the coordinate acceleration is positive all the way from the horizon to spatial infinity.}\label{F:sch-sch-null-ddot-r}
\end{figure}

The radial coordinate velocity and radial coordinate acceleration for timelike geodesics are plotted as shown in figures \ref{F:sch-sch-dot-r} and  \ref{F:sch-sch-ddot-r}.
For null geodesics see figures \ref{F:sch-sch-null-dot-r} and  \ref{F:sch-sch-null-ddot-r}.
\clearpage


\clearpage
\subsection{Isotropic coordinates}
The Schwarzschild geometry in isotropic coordinates is described by \cite{Wheeler}
\begin{equation}
\d s^{2} = -\frac{\left(1 - \frac{m}{2r}\right)^{2}}{\left(1+\frac{m}{2r}\right)^{2}}\d t^{2} + \left(1 + \frac{m}{2r}\right)^{4}\left[\d r^{2} + r^{2}\left(\d\theta^{2} + \sin^{2}\theta \, \d\phi^{2}\right)\right].
\end{equation}
Compared to Schwarzschild curvature coordinates, only the meaning of the $r$ coordinate has changed. Indeed\footnote{In these isotropic coordinates the horizon is now at $r = {m\over2}$.}
\begin{equation}
r_\mathrm{Schwarzschild} =  r_\mathrm{isotropic} \left(1+\frac{m}{2r_\mathrm{isotropic}}\right)^{2}.
\end{equation}

The Christoffel symbols are easily calculated and the geodesic equation becomes
\begin{equation}
\ddot r =  - {m\over r^2} {(1-{m\over2r})\over(1+{m\over2r})^7} +3\, {m\over r^2}\, {(1-{m\over 6r})\over (1-{m\over2 r})(1+{m\over 2r})}   \,\dot r^2.
\end{equation}
This can also be recast as
\begin{equation}
\ddot r =  - {m\over r^2} \left\{ {(1-{m\over2r})\over(1+{m\over2r})^7} -3\, {(1-{m\over 6r})\over (1-{m\over2 r})(1+{m\over 2r})}   \,\dot r^2 \right\}.
\end{equation}
This already demonstrates, (working in terms of $r$ and $\dot r$), that $\ddot r$ \emph{changes sign} at the critical coordinate velocities
\begin{equation}
(\dot r)_*^2 = {1\over3} {\left(1-{m\over2 r}\right)^2\over\left(1+{m\over 2r}\right)^6(1-{m\over 6r})}
\end{equation}
At large $r$, (ie weak fields), this sign flip takes place at $\dot r^2 \approx 1/3$;  this is mildly relativistic but certainly not ultra-relativistic. (In the weak-field limit the Schwarzschild curvature coordinates and the isotropic coordinates asymptotically approach each other.)

From the Killing conservation equation, since the metric in isotropic coordinates is diagonal,  we deduce
\begin{equation}
\dot r =
\pm {\sqrt{ (\epsilon^2+g_{tt})(-g_{tt}g_{rr}) } \over \epsilon g_{rr}} =
\pm {\sqrt{ (1+\epsilon^{-2}g_{tt})(-g_{tt}g_{rr}) } \over g_{rr}}.
\label{E:quadratic}
\end{equation}
This implies
\begin{equation}
\dot r = \pm {1\over\epsilon} \sqrt{\epsilon^2 - \left(1-{m\over2r}\over1+{m\over2r}\right)^2}
\;
{\left(1-{m\over2r}\right)\over \left(1+{m\over2r}\right)^3}.
\end{equation}
Note that at the horizon, now located at $r=m/2$, we again have $\dot r \to 0$, while at spatial infinity we agin see $\dot r \to \sqrt{1-\epsilon^{-2}}$ at fixed energy.
Combining these results, for a dropped particle (of fixed energy $\epsilon$) we have
\begin{equation}
\ddot r =- {m\over r^2} {(1-{m\over2r})\over(1+{m\over2r})^7}
\left( 1 - {3(1-{m\over6r})\over\epsilon^2} \left[ \epsilon^2 - \left(1-{m\over2r}\over1+{m\over2r}\right)^2 \right]  \right).
\end{equation}
Note that the coordinate acceleration $\ddot r$ goes through zero and, (apart from the trivial zero at $r_*=m/2$), changes \emph{sign}
at the critical values $r_*$ of $r$ given by solving the cubic equation
\begin{equation}
r_*:\qquad
 1 - {3(1-{m\over6r})\over\epsilon^2} \left[ \epsilon^2 - \left(1-{m\over2r}\over1+{m\over2r}\right)^2 \right]  =0.
\end{equation}
For a particle dropped at rest from spatial infinity ($\epsilon=1$) this simplifies to\footnote{Though not entirely obvious, it is easy to check that at large distances $\dot r\to \sqrt{2m/r}$, as expected from the Newtonian limit. It is more obvious that at large distances $\ddot r \to - m/r^2$. In isotropic coordinates, the ISCO is at $\left({5\over2}+\sqrt{6}\right)m$, which is not where $\ddot r \to 0$; that these two locations coincided in curvature coordinates is merely a coincidence.}
\begin{equation}
\dot r = - \sqrt{1 - \left(1-{m\over2r}\over1+{m\over2r}\right)^2}
\;
{\left(1-{m\over2r}\right)\over \left(1+{m\over2r}\right)^3};
\end{equation}
\begin{equation}
\ddot r =- {m\over r^2}\;  {(1-{m\over2r})\over(1+{m\over2r})^7}
\left( 1 - {3\left(1-{m\over6r}\right)} \left[1 - \left(1-{m\over2r}\over1+{m\over2r}\right)^2 \right]  \right);
\end{equation}
with the (non-trivial) zeros of coordinate acceleration determined by
\begin{equation}
 1 - {3\left(1-{m\over6r}\right)} \left[ 1 - \left(1-{m\over2r}\over1+{m\over2r}\right)^2 \right]  =0;
 \qquad
 r_* =  (5\pm 2\sqrt{5}) {m\over2}.
\end{equation}
For null geodesics ($\epsilon\to\infty$) we have
\begin{equation}
\dot r = -{\left(1-{m\over2r}\right)\over \left(1+{m\over2r}\right)^3};
\qquad
\ddot r =  {2m\over r^2} {(1-{m\over2r})( 1- {m\over4r})\over(1+{m\over2r})^7}.
\end{equation}

\begin{figure}[h]
\centerline{\psfig{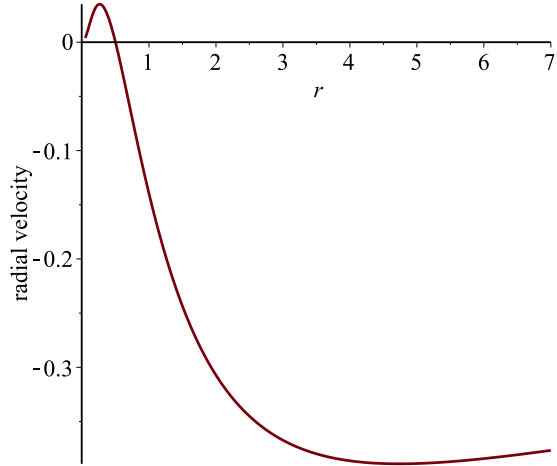}}
\vspace*{1pt}
\caption{Behaviour of $\dot{r}$ in the Schwarzschild geometry using isotropic coordinates for $m = 1$ and $\epsilon = 1$. Note the coordinate velocity is negative all the way from the horizon (now at $m/2$) to spatial infinity.}\label{F:sch-iso-dot-r}
\end{figure}

\begin{figure}[h]
\centerline{\psfig{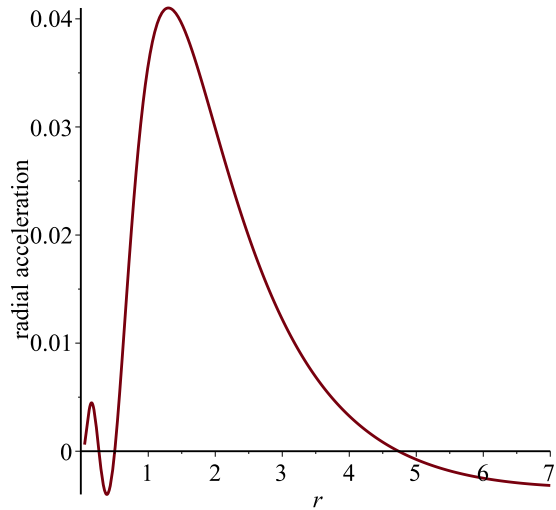}}
\vspace*{1pt}
\caption{Behaviour of $\ddot{r}$ in the Schwarzschild geometry using isotropic coordinates for $m = 1$ and $\epsilon = 1$.
Note that the curve crosses the $r$ axis both at $r=1/2$ and $r={5+2\sqrt{5}\over2} \approx 4.736067977$; there is a third unphysical root at  $r={5-2\sqrt{5}\over2} \approx 0.263932023$.}\label{F:sch-iso-ddot-r}
\end{figure}

\begin{figure}[h]
\centerline{\psfig{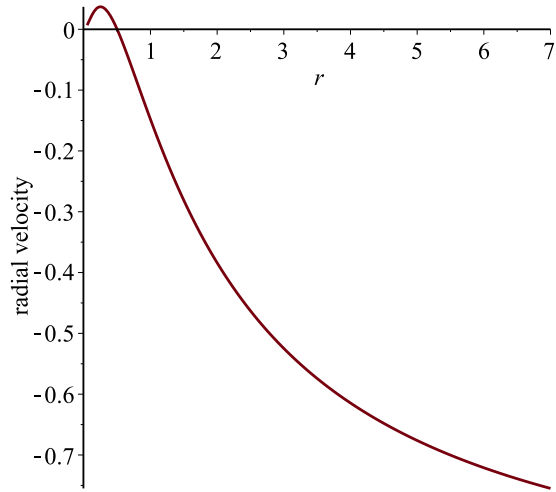}}
\vspace*{1pt}
\caption{Behaviour of $\dot{r}$ for null geodesics in the Schwarzschild geometry using isotropic coordinates for $m = 1$ and $\epsilon\to\infty$.
Note the coordinate velocity is negative all the way from the horizon (now at $m/2$) to spatial infinity.}\label{F:sch-iso-null-dot-r}
\end{figure}

\begin{figure}[h]
\centerline{\psfig{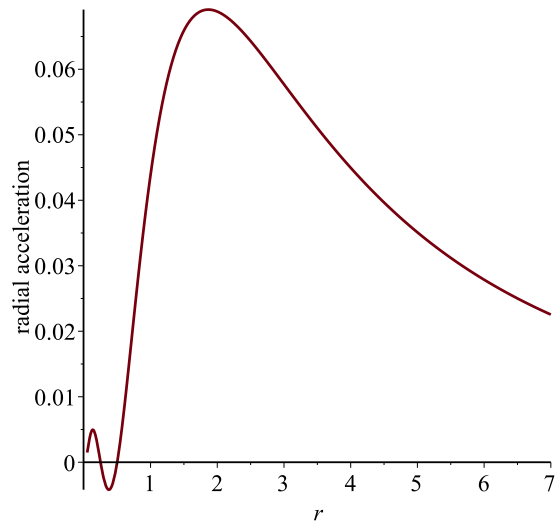}}
\vspace*{1pt}
\caption{Behaviour of $\ddot{r}$ for null geodesics in the Schwarzschild geometry using isotropic coordinates for $m = 1$ and $\epsilon\to\infty$. Note horizon is now at $m/2$; there is an extra zero at $m/4$.
Note the coordinate acceleration is positive all the way from the horizon to spatial infinity.}\label{F:sch-iso-null-ddot-r}
\end{figure}

The radial coordinate velocity and radial coordinate acceleration for timelike geodesics are plotted as shown in figures \ref{F:sch-iso-dot-r} and  \ref{F:sch-iso-ddot-r}.
For null geodesics see figures \ref{F:sch-iso-null-dot-r} and  \ref{F:sch-iso-null-ddot-r}.
Note the similarities, \emph{and differences}, compared to what we saw for Schwarzschild curvature coordinates.



\clearpage
\subsection{Painleve--Gullstrand coordinates}
The Schwarzschild geometry in Painleve--Gullstrand coordinates is described by \cite{Painleve, Gullstrand, unexpected,Acoustic,Analogue}
\begin{equation}
\d s^{2} = -\left(1 - \frac{2m}{r}\right)\d t_{PG}^{2} + 2\sqrt{\frac{2m}{r}}\,\d t_{PG}\,\d r + \d r^{2} + r^{2}\left(\d\theta^{2} + \sin^{2}\theta\, \d\phi^{2}\right),
\end{equation}
where the  Painleve--Gullstrand time coordinate is given in terms of the Schwarzschild time coordinate by
\begin{equation}
t_{PG} = t_{Schwarzschild} - 2m\left[2\sqrt{\frac{r}{2m}} - \ln\left(\frac{1 + \sqrt{2m/r}}{1 - \sqrt{2m/r}}\right)\right].  
\end{equation}
Note in particular that $g_{tt} \, g_{rr}-g_{tr}^2 = -1$.

It is easy to calculate the Christoffell symbols and verify that in these coordinates the radial geodesic equation becomes
\begin{equation}
\ddot r = -{m\over r^2}\left(1-{2m\over r}\right) +\frac{3m}{r^{2}}\sqrt{\frac{2m}{r}}\,\dot{r} + \frac{3m}{r^{2}}\,\dot{r}^{2} + \frac{\sqrt{2m/r}}{2r}\,\dot{r}^{3}.
\end{equation}
We can rewrite this as
\begin{equation}
\ddot r = -{m\over r^2}\left\{ \left(1-{2m\over r}\right) -3\sqrt{\frac{2m}{r}}\,\dot{r} - 3\dot{r}^{2}\right\} + \frac{\sqrt{2m/r}}{2r}\,\dot{r}^{3}.
\end{equation}

Viewed as a function of $\dot r$, this flips sign at the critical coordinate velocity\footnote{
 This actually implies that $\ddot r$ factorizes as follows: $\ddot r = \left(\dot r - \left[ \sqrt[6]{2m\over r} - \sqrt{2m\over r}\right]\right) \times \hbox{(quadratic in $\dot r$)}$, where the quadratic has no real zeroes. Unfortunately the specific form of the quadratic is too messy to be illuminating.}
\begin{equation}
\left(\dot r\right)_* =  \sqrt[6]{2m\over r} - \sqrt{2m\over r},
\end{equation}
which is always positive outside the horizon.

In view of the fact that in these coordinates $-g_{tt} \, g_{rr}+g_{tr}^2 = 1$, the general Killing-induced result for the coordinate velocity simplifies to
\begin{equation}
\dot r =  {-g_{tr}(\epsilon^2+g_{tt})
\pm \epsilon \sqrt{ (\epsilon^2+g_{tt}) }
\over \epsilon^2 g_{rr} + g_{tr}^2}.
\label{E:quadratic-5}
\end{equation}
Thence
\begin{equation}
\dot r =  {-\sqrt{2m/r}(\epsilon^2-1+2m/r)
\pm \epsilon \sqrt{ (\epsilon^2-1+2m/r) }
\over \epsilon^2 + 2m/r}.
\label{E:quadratic-6}
\end{equation}

This can also be written as
\begin{equation}
\dot r = -\sqrt{\frac{2m}{r}} + {\sqrt{2m/r} \pm \epsilon \sqrt{ (\epsilon^2-1+2m/r) }
\over \epsilon^2 + 2m/r}.
\end{equation}

At the horizon, $r=2m$, we have
\begin{equation}
(\dot r)_H \in \left\{ 0,  {-2\epsilon^2\over\epsilon^2+1} \right\}.
\label{E:quadratic-6}
\end{equation}
Therefore, we see that the ingoing geodesic crosses the horizon with finite coordinate velocity\footnote{Note that $|\dot r|$ can easily exceed unity; this is just a coordinate speed, not a physical speed.}
\begin{equation}
(\dot r)_H = -{2\epsilon^2\over\epsilon^2+1} = -{2\over1+\epsilon^{-2}} \in (-2,0),
\end{equation}
 while the outgoing geodesic crosses the horizon with coordinate velocity zero. This makes it clear that for a dropped particle we should take the negative root in $\dot r$ so that:
\begin{equation}
\dot r =  -\sqrt{ 1- \epsilon^{-2}(1-2m/r) } \;\;
  {1+\sqrt{2m/r} \sqrt{ 1- \epsilon^{-2}(1-2m/r) } \over 1+ \epsilon^{-2} (2m/r)}.
\label{E:quadratic-6}
\end{equation}

Combining these results, for a dropped particle (fixed energy $\epsilon$) we have
\begin{eqnarray}
\ddot r &=& -{m\over r^2}\left(1-{2m\over r}\right)\nonumber\\
        &&  + \frac{3m}{r^{2}}\sqrt{\frac{2m}{r}}\,\left[{-\sqrt{2m/r}(\epsilon^2-1+2m/r)- \epsilon \sqrt{ (\epsilon^2-1+2m/r) }\over \epsilon^2 + 2m/r}\right]\nonumber\\
        &&  + \frac{3m}{r^{2}}\,\left[{-\sqrt{2m/r}(\epsilon^2-1+2m/r)- \epsilon \sqrt{ (\epsilon^2-1+2m/r) }\over \epsilon^2 + 2m/r}\right]^{2}\nonumber\\
        && + \frac{\sqrt{2m/r}}{2r}\,\left[{-\sqrt{2m/r}(\epsilon^2-1+2m/r)- \epsilon \sqrt{ (\epsilon^2-1+2m/r) }\over \epsilon^2 + 2m/r}\right]^{3}.
\end{eqnarray}

For a timelike particle dropped at rest from spatial infinity ($\epsilon$ = 1) this simplifies quite drastically to yield\footnote{The fact that in this particular situation the Painleve--Gullstrand coordinate system exactly reproduces the Newtonian result is one of the many reasons that
the Painleve--Gullstrand coordinate system is so useful.}
\begin{equation}
\dot r =  - \sqrt{2m\over r};  \qquad \ddot r = - {m\over r^2}.
\label{E:quadratic-6}
\end{equation}
Note that for $\epsilon=1$  the (ingoing) coordinate acceleration $\ddot{r}$ is extremely simple,  and always negative. In fact the coordinate acceleration is finite at horizon crossing $(\ddot r)_H = -1/(4m)$.

For an infalling light-like particle ($\epsilon\to\infty$) this again simplifies quite drastically to yield
\begin{equation}
\dot r =  - 1 - \sqrt{2m\over r};  \qquad
\ddot r = - {1\over2r } \sqrt{2m\over r} \left( 1 + \sqrt{2m\over r}\right)
\label{E:quadratic-6}
\end{equation}
Note that for $\epsilon\to\infty$  the (ingoing) coordinate acceleration $\ddot{r}$ is relatively simple,  and always negative. In fact the coordinate acceleration is finite at horizon crossing $(\ddot r)_H = -1/(2m)$.
(The situation for Painleve--Gullstrand coordinates is ultimately so simple that graphs are not needed.)

\subsection{Kerr--Schild coordinates}
The Schwarzschild geometry in Kerr--Schild coordinates is described by \cite{Stephani, Kerr}
\begin{eqnarray}
\d s^{2} &=& -\d t^{2} + \d r^{2} + r^{2}\left(\d\theta^{2} + \sin^{2}\theta\, \d\phi^{2}\right) + \frac{2m}{r}(\d t + \d r)^{2}.
\end{eqnarray}
That is
\begin{eqnarray}
\d s^{2}
       &=& -\left(1 - \frac{2m}{r}\right)\d t^{2} + \frac{4m}{r}\d t\d r + \left(1 + \frac{2m}{r}\right)\d r^{2} + r^{2}\left(\d\theta^{2} + \sin^{2}\theta\, \d\phi^{2}\right).
\end{eqnarray}
Note in particular that $g_{tt} \, g_{rr}-g_{tr}^2 = -1$.

The Christoffel symbols are easily calculated and in these coordinates the radial geodesic equation becomes
\begin{equation}
\ddot r = -{m\over r^2}\left(1-{2m\over r}\right) +\frac{6m^2}{r^{3}}\,\dot{r} + \frac{3m}{r^{2}}\,\left(1+{2m\over r}\right)\dot{r}^{2} + {2m\over r^2} \left(1+{m\over r}\right)\dot{r}^{3}.
\end{equation}
It may be better to rewrite this as follows:
\begin{equation}
\ddot r = -{m\over r^2}\left\{ \left(1-{2m\over r}\right) -\frac{6m}{r}\,\dot{r} - 3 \left(1+{2m\over r}\right)\dot{r}^{2} -2  \left(1+{m\over r}\right)\dot{r}^{3}
\right\}.
\end{equation}
This factorizes
\begin{equation}
\ddot r = -{m\over r^2} \; (1+\dot r)^2 \left\{ \left(1-{2m\over r}\right) - 2\left(1+{m\over r}\right)\dot{r}
\right\}.
\end{equation}
As a function of $\dot r$ we see that $\ddot r$ flips sign at the critical coordinate velocity
\begin{equation}
\left(\dot r\right)_*  = {1-2m/r\over2(1+m/r)},
\end{equation}
This is always positive, and less than 1/2,  outside the horizon.

In view of the fact that in these coordinates $-g_{tt} \, g_{rr}+g_{tr}^2 = 1$, the general Killing-induced result for the coordinate velocity simplifies to
\begin{equation}
\dot r =  \pm \sqrt{ (1+\epsilon^{-2}g_{tt}) } \;\; {1\mp g_{tr}\sqrt{1+\epsilon^{-2}g_{tt}}
\over  g_{rr} + \epsilon^{-2} g_{tr}^2}.
\label{E:quadratic-5}
\end{equation}
Thence
\begin{equation}
\dot r =
\pm \sqrt{ 1-\epsilon^{-2}(1-2m/r)}\;\;  { 1 \mp (2m/r)\sqrt{1-\epsilon^{-2}(1-2m/r)}
\over (1 + 2m/r) +\epsilon^{-2}(2m/r)^2 }.
\label{E:quadratic-6}
\end{equation}

At the horizon
\begin{equation}
(\dot r)_H \in \left\{ 0, -{2\epsilon^2\over2\epsilon^2+1} \right\}.
\end{equation}
Therefore, the ingoing geodesic crosses the horizon with finite coordinate velocity
\begin{equation}
(\dot r)_H = -{2\epsilon^2\over2\epsilon^2+1} \in \left(-1,-{2\over3}\right),
\end{equation}
while the outgoing geodesic crosses the horizon with coordinate velocity zero. This makes it clear that for a dropped particle we should take the negative root in $\dot r$.

Combining these results, for a dropped (ingoing) particle (fixed energy $\epsilon$) we have
\begin{equation}
\dot r =
- \sqrt{ 1-\epsilon^{-2}(1-2m/r)}\;\;  { 1 + (2m/r)\sqrt{1-\epsilon^{-2}(1-2m/r)}
\over (1 + 2m/r) +\epsilon^{-2}(2m/r)^2 }.
\label{E:quadratic-6}
\end{equation}
For unbound particles ($\epsilon\geq 1$) this is negative real everywhere, both outside and inside the horizon; in fact all the way down to $r=0$. For the coordinate acceleration
\begin{eqnarray}
\ddot r &=& -{m\over r^2}\left\{ \left(1-{2m\over r}\right)\right.\nonumber\\
        &&  -\frac{6m}{r}\,\left[{-(2m/r)(\epsilon^2-1+2m/r) - \epsilon \sqrt{ (\epsilon^2-1+2m/r) }\over \epsilon^2(1 + 2m/r) +(2m/r)^2 }\right]\nonumber\\
        &&  - 3 \left(1+{2m\over r}\right)\left[{-(2m/r)(\epsilon^2-1+2m/r) - \epsilon \sqrt{ (\epsilon^2-1+2m/r) }\over \epsilon^2(1 + 2m/r) +(2m/r)^2 }\right]^{2}\nonumber\\
        &&  \left.-2  \left(1+{m\over r}\right)\left[{-(2m/r)(\epsilon^2-1+2m/r) - \epsilon \sqrt{ (\epsilon^2-1+2m/r) }\over \epsilon^2(1 + 2m/r) +(2m/r)^2 }\right]^{3}
\right\}.
\end{eqnarray}

At the horizon, $r=2m$, we have
\begin{equation}
(\ddot r)_H = - {3\over2} {\epsilon^2\over (2\epsilon^2+1)^3 m} ,
\end{equation}
a finite inward coordinate acceleration.

For a timelike particle dropped at rest from spatial infinity ($\epsilon$ = 1) this simplifies to\footnote{Note that asymptotically $\dot r \to -\sqrt{2m/r}$, as expected from the Newtonian limit.}
\begin{equation}
\dot r = -\left[{(2m/r)^{2} + \sqrt{ 2m/r }\over (1 + 2m/r) +(2m/r)^2 }\right]
=
-\sqrt{2m\over r}  \left[{1+(2m/r)^{3/2} \over 1 + 2m/r +(2m/r)^2 }\right].
\end{equation}
and
\begin{eqnarray}
\ddot r &=& -{m\over r^2}\left\{ \left(1-{2m\over r}\right) +\frac{6m}{r}\,\left[{(2m/r)^{2} + \sqrt{ (2m/r) }\over (1 + 2m/r) +(2m/r)^2 }\right]\right.\nonumber\\
        &&  - 3 \left(1+{2m\over r}\right)\left[{(2m/r)^{2} + \sqrt{ (2m/r) }\over (1 + 2m/r) +(2m/r)^2 }\right]^{2}\nonumber\\
        &&  \left. +2  \left(1+{m\over r}\right)\left[{(2m/r)^{2} + \sqrt{ (2m/r) }\over (1 + 2m/r) +(2m/r)^2 }\right]^{3}
\right\}.
\end{eqnarray}

\clearpage
Note that the coordinate acceleration $\ddot{r}$, and the coordinate velocity $\dot r$, are both always negative. 
We can also factorize this as:
\begin{eqnarray}
\ddot r &=& -{m\over r^2}\left[1 - \sqrt{2m\over r}  \left\{{1+(2m/r)^{3/2} \over 1 + 2m/r +(2m/r)^2 }\right\}\right]^{2}\nonumber\\
        &&  \times \left\{ \left(1-{2m\over r}\right) + 2\sqrt{\frac{2m}{r}}\left(1+{m\over r}\right)\left[{1+(2m/r)^{3/2} \over 1 + 2m/r +(2m/r)^2 }\right]
\right\}.
\end{eqnarray}

\begin{figure}[!h]
\centerline{\psfig{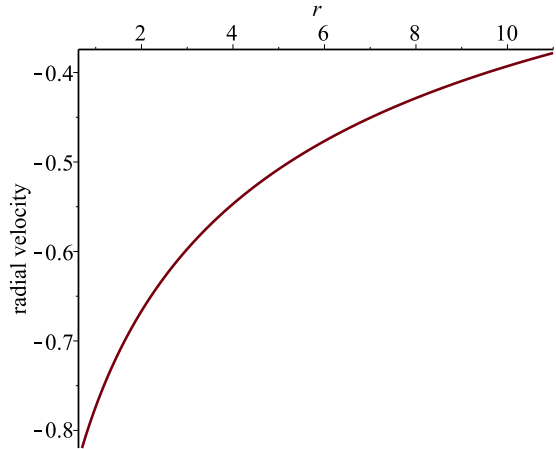}}
\vspace*{1pt}
\caption{Behaviour of $\dot{r}$ for the Schwarzschild geometry in Kerr--Schild coordinates for $m = 1$ and $\epsilon = 1$.
Note the coordinate velocity at the horizon is $-2/3$, and that the coordinate velocity remains negative between the horizon and spatial infinity.}\label{F:sch-ks-v}
\end{figure}

\begin{figure}[!htbp]
\centerline{\psfig{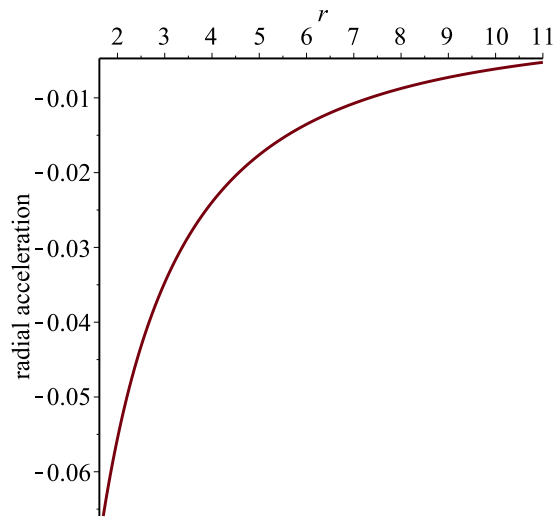}}
\vspace*{1pt}
\caption{Behaviour of $\ddot{r}$ for the Schwarzschild geometry in Kerr--Schild coordinates for $m = 1$ and $\epsilon = 1$.
Note the coordinate acceleration remains negative between the horizon and spatial infinity.
}\label{F:sch-ks-a}
\end{figure}

\noindent We can see that when approaching the horizon, at fixed $\epsilon=1$, we have
\begin{equation}
(\dot r)_H = -{2\over3}; \qquad (\ddot r)_H = -{1\over 18m}.
\end{equation}
The radial coordinate velocity and radial coordinate acceleration are plotted as shown in figures \ref{F:sch-ks-v} and \ref{F:sch-ks-a} respectively.

Finally, note that for a light-like particle ($\epsilon\to\infty$) in Kerr--Schild coordinates we have the very drastic simplification\footnote{So in Kerr--Schild coordinates ingoing photons happen to have coordinate acceleration zero. This is one reason Kerr--Schild coordinates are popular.}
\begin{equation}
\dot r = -1; \qquad \ddot r = 0.
\end{equation}
(For this particular case a figure would be entirely superfluous.)


\clearpage
\section{Conclusions}
Now that we have seen some specific examples of what happens to near-horizon geodesics in various coordinate systems, let us attempt to draw some general inferences.
{While the specific computations in this article have been carried out for the Schwarzschild geometry, this is known to be a good approximation for slowly rotating astrophysical black holes, and for numerical simulations of black holes, and even for semi-classical black holes in the Unruh quantum vacuum---so the overall conclusions are generic to a wide class of physically and observationally interesting black holes.
} 

The most obvious conclusion we can draw is that the coordinate velocity, and coordinate acceleration, are (quite naturally) extremely coordinate dependent, and that no general physical conclusions can be drawn from the magnitude of the coordinate velocity, ($\dot r$ can easily exceed unity), or the \emph{sign} of the coordinate acceleration $\ddot r$. Claims that gravity in general relativity is ``repulsive'' at high speeds and/or near the horizon are at best disingenuous --- they are merely misinterpretations of coordinate artefacts. For a fixed spacetime, by suitably choosing the coordinate system we can easily make $(\ddot r)_H=0$ or $(\ddot r)_H=\hbox{(finite negative)}$ at horizon crossing.
For a fixed spacetime, by suitably choosing the coordinate system we can easily make the coordinate acceleration $\ddot r$ \emph{either positive or negative} just prior to horizon crossing. Indiscriminately mixing general relativistic and Newtonian concepts is dangerous and misleading.

The major distinction we have seen in the specific examples we explored was in the difference between horizon-penetrating and horizon-non-penetrating coordinates. There are good physical and mathematical reasons for this. In horizon-non-penetrating coordinates geodesics (essentially by definition) pile up at the horizon and do not cross it --- in coordinates of this type $|\dot r|$ first increases as one falls inwards, but then has to go to zero at the horizon. This implies that $|\dot r|$ must have a maximum where $\partial_r \dot r =0$ and hence $\ddot r =0$.  Thence regions where the coordinate acceleration is positive $\ddot r >0$ are \emph{unavoidable} in horizon-non-penetrating coordinates.
In contrast horizon-penetrating coordinates are much better behaved when studying near horizon physics, with the coordinate velocity and coordinate acceleration being non-zero and finite at horizon crossing.

\enlargethispage{10pt}

\section*{Acknowledgement}
This project was funded by the Ratchadapisek Sompoch Endowment Fund, Chulalongkorn University (Sci-Super 2014-032), by a grant for the professional development of new academic staff from the Ratchadapisek Somphot Fund at Chulalongkorn University, by the Thailand Research Fund (TRF), and by the Office of the Higher Education Commission (OHEC), Faculty of Science, Chulalongkorn University (RSA5980038). PB was additionally supported by a scholarship from the Royal Government of Thailand. TN was also additionally supported by a scholarship from the Development and Promotion of Science and Technology talent project (DPST). MV was supported by the Marsden Fund, via a grant administered by the Royal Society of New Zealand.


\end{document}